\documentclass[12pt]{article}

\usepackage{cite,hyperref,amsfonts,amsmath,amssymb,bm,graphicx,subfigure,wasysym,a4wide}

\numberwithin{equation}{section}

\begin{document}

\title{\textbf{Influence of the turbulent motion on the chiral magnetic effect in the early Universe}}

\author{Maxim Dvornikov$^{a,b}$\thanks{maxdvo@izmiran.ru},\ 
Victor B. Semikoz$^{a}$\thanks{semikoz@yandex.ru}
\\
$^{a}$\small{\ Pushkov Institute of Terrestrial Magnetism, Ionosphere} \\
\small{and Radiowave Propagation (IZMIRAN),} \\
\small{108840 Moscow, Troitsk, Russia;} \\
$^{b}$\small{\ Physics Faculty, National Research Tomsk State University,} \\
\small{36 Lenin Avenue, 634050 Tomsk, Russia}}

\date{}

\maketitle

\begin{abstract}
We study the magnetohydrodynamics of relativistic plasmas accounting for the chiral magnetic effect (CME). To take into account the evolution of the plasma velocity, obeying the Navier-Stokes equation, we approximate it by the Lorentz force accompanied by the phenomenological drag time parameter. On the basis of this ansatz, we obtain the contributions of both the turbulence effects, resulting from the dynamo term, and the magnetic field instability, caused by the CME, to the
evolution of the magnetic field governed by the modified Faraday equation. In this way, we explore the evolution of the magnetic field energy and the magnetic helicity density spectra in the early Universe plasma. We find that the right-left electron asymmetry is enhanced by the turbulent plasma motion in a strong seed magnetic field compared to the pure the CME case studied earlier for the hot Universe plasma in the same broken phase.
\end{abstract}

\maketitle

\section{Introduction}

Magnetic fields are important for various physical processes, including the cosmic rays propagation, influence on the stellar and solar activities, etc. However, the origin of cosmic magnetic fields is still an open problem in astrophysics and cosmology~\cite{Brandenburg:2012jz,Grasso:2000wj,Durrer:2013pga}. It remains unclear whether these magnetic fields were first created by battery effects in protogalaxies and then amplified by a dynamo action up to their present-day strengths, or if seed fields for a dynamo action  originated in magnetic fields which seem to have existed in the early Univese before the recombination. The first observational indications of the presence of a cosmological magnetic field (CMF) in the intergalactic medium~\cite{Neronov:1900zz} still do not preclude the first possibility. However, they strongly support the latter option.

The origin of the CMF, as well as  a primeval chiral asymmetry $\mu_5=(\mu_{e\mathrm{R}} - \mu_{e\mathrm{L}})/2$, where $\mu_{e\mathrm{R}}$ and $\mu_{e\mathrm{L}}$ are the right-  and left-electron chemical potentials, can be traced from the lepto- and baryogenesis in primordial hypermagnetic fields existing in the symmetric phase of the Universe before the electroweak phase transition (EWPT); e.g., in the model with a nonzero initial right-electron asymmetry $\sim \mu_{e\mathrm{R}}\neq 0$~\cite{Giovannini:1997eg,Dvornikov:2011ey,Semikoz:2012ka,Dvornikov:2012rk}. The important issue in such a scenario is a nonzero difference of lepton numbers, $L_{e\mathrm{R}} - L_{e\mathrm{L}}\neq 0$ at the EWPT time \cite{Dvornikov:2011ey}, that can be used as a possible starting value for the chiral anomaly which provides the evolution of Maxwellian fields down to the temperatures $T\sim 100~{\rm MeV}\gg m_e$ where $m_e$ is the electron mass. 

There are different ways to estimate the importance of the advection (dynamo) term $\nabla \times (\mathbf{v}\times \mathbf{B})$ in the Faraday equation describing CMF. In Ref.~\cite{Boyarsky:2011uy}, accounting for the chiral magnetic effect (CME)~\cite{Vilenkin} given by $\mu_5\neq 0$, one neglects the velocity field $\mathbf{v}$ completely. In Ref.~\cite{Tashiro:2012mf},
considering a negligible backreaction of the magnetic field on the fluid velocity, it is shown that the advection term is unimportant. A different suggestion on the plasma velocity $\mathbf{v}$ and the advection term is put forward in Ref.~\cite{Campanelli:2007tc}, assuming that the backreaction of a strong magnetic field on a fluid is important, cf. 
Refs.~\cite{Sigl:2002kt,Subramanian:1999mr,Cornwall:1997ms}.

There is also an interesting discussion in literature~\cite{Pavlovic:2016mxq} on the inverse cascade induced by CME. However, in the present work, we do not deal with this topic  trying to illuminate other aspects of anomalous MHD in the presence of a fluid turbulence in chiral plasma.

In this paper, we revisit the idea of Ref.~\cite{Campanelli:2007tc}. In Sec.~\ref{sec:SIMPL}, we simplify the Navier-Stokes equation substituting for the velocity field $\mathbf{v}$ entering the dynamo term in the Faraday equation its expression through  the Lorentz force as suggested in Ref.~\cite{Campanelli:2007tc}. Then, in Sec.~\ref{sec:KINETIC}, we derive the system of the magnetohydrodynamic (MHD) equations describing the evolution of the magnetic field energy and magnetic helicity density spectra. Making some assumptions in Sec.~\ref{sec:ANALSOL}, we represent the kinetic equations in the form of the integral equations. In Sec.~\ref{sec:NUMSOL}, we solve the nonlinear kinetic equations numerically. Finally, in Sec.~\ref{sec:DISC}, we discuss our results comparing them with those obtained earlier. Some details of the derivation of kinetic equations for the spectra are provided in Appendix~\ref{sec:KINDER}. In the following we use the natural units, in which $\hbar=c=1$. 

\section{Simplification of the set of MHD equations\label{sec:SIMPL}}

In the present work, we extend the approach developed in Ref.~\cite{Boyarsky:2011uy} considering a hot plasma of the early Universe in the broken phase after EWPT at the relativistic temperatures $10\,\mathrm{MeV}<T< T_\mathrm{EWPT}\simeq 100\,\mathrm{GeV}$. For the system obeying the equation of state $P=\rho/3$ or $P + \rho=4\rho/3$,
the MHD equations in the radiation-matter single fluid approximation are
\begin{eqnarray}\label{MHD}
  &&\partial_t\rho + \frac{4}{3}\nabla\cdot(\rho \mathbf{v})=0,
  \\
  &&\frac{4}{3}\partial_t(\rho\mathbf{v}) -
  \frac{4}{3}\rho\mathbf{v}\times(\nabla\times \mathbf{v}) =
  (\mathbf{J}\times \mathbf{B}) - \nabla P +
  \frac{4}{3}\rho\nu\nabla^2\mathbf{v}, 
  \label{eq:NS}
  \\
  && \mathbf{J}= (\nabla\times \mathbf{B}),
  \\
  &&\partial_t\mathbf{B} = -\nabla\times \mathbf{E},
  \label{eq:Bianchi} 
\end{eqnarray}
where $\rho$ is the energy density of the fluid, $P$ is the pressure, and $\mathbf{E}$ is the electric field.

Using Eq.~\eqref{eq:Bianchi} as well as accounting for the total electric current $\mathbf{J}=\mathbf{J}_\mathrm{Ohm} + \mathbf{J}_\mathrm{CME}$ and the Ohm law $\mathbf{J}_\mathrm{Ohm}=\sigma_\mathrm{cond}[\mathbf{E} + (\mathbf{v}\times \mathbf{B})]$,  with the anomalous current $\mathbf{J}_\mathrm{CME}$ directed along the magnetic field, $\mathbf{J}_\mathrm{CME}=(2\alpha_\mathrm{em}\mu_5/\pi)\mathbf{B}$, one finds the Faraday equation modified due to CME,
\begin{equation}\label{Faraday}
  \partial_t\mathbf{B} = \nabla \times (\mathbf{v}\times\mathbf{B}) +   
  \eta_m\nabla^2\mathbf{B} +
  \frac{2\alpha_\mathrm{em}\mu_5}{\pi \sigma_\mathrm{cond}}\nabla\times \mathbf{B}.
\end{equation}
Here $\alpha_\mathrm{em}=e^2/4\pi \approx 1/137$ is the fine structure constant, $\eta_m=\sigma_\mathrm{cond}^{-1}$ is the magnetic diffusion coefficient, $\sigma_\mathrm{cond}=\sigma_cT$ is the hot plasma conductivity, and $\sigma_c\simeq 100$.
Using the Faraday equation~(\ref{Faraday}) without the dynamo term $\nabla\times (\mathbf{v}\times \mathbf{B})$ completed by the chiral imbalance evolution equation [see Eq.~(\ref{mu5}) below], the evolution of the binary products, such  as the magnetic energy density $E_\mathrm{B}\sim B^2$ and the magnetic helicity density $H_\mathrm{B} = V^{-1}\smallint \mathrm{d}^3 x ({\bf A} \cdot \mathbf{B})$, is studied in Ref.~\cite{Boyarsky:2011uy}. In the present work we analyze the importance of the dynamo term neglected in Ref.~\cite{Boyarsky:2011uy} and interpreted in Ref.~\cite{Campanelli:2007tc} as the turbulent fluid contributions to the evolution of the magnetic energy and helicity density spectra.

Since the fluid velocity $\mathbf{v}$ should obey the Navier-Stokes equation~\eqref{eq:NS}, which is rather difficult to solve, we use as in Ref.~\cite{Campanelli:2007tc}
the following approximation instead:
\begin{equation}\label{short}
  \frac{\partial \mathbf{v}}{\partial t}=
  \frac{1}{P+ \rho}(\mathbf{J}\times \mathbf{B}),
\end{equation}
where we drop all the gradients referring to the matter variables, including the pressure and the kinematic viscosity ($\sim \nu$) terms as well as the nonlinear velocity term. Thus, only the Lorentz-force term is retained on the right-hand side of Navier-Stokes equation~\eqref{eq:NS}, $\mathbf{F}_\mathrm{L}\sim (\mathbf{J}\times \mathbf{B})$. Then we simplify Eq.~(\ref{short}) representing it as
\begin{equation}\label{velocity}
  \mathbf{v}=
  \frac{\tau_d}{P+ \rho}
  (\mathbf{J}\times \mathbf{B}),
\end{equation}
where $\tau_d=l_\mathrm{free}\approx 1/\alpha_\mathrm{em}^2T$ is the correlation (drag) time. The drag time is the average time  of the Coulomb scattering in a hot plasma~\cite{fn1}, which is much greater than the period of the Larmor rotation. It means that the charged fluid can be accelerated by the Lorentz force until it interacts with other particles in the background. 

The physical meaning of our choice for the drag time $\tau_d=l_\mathrm{free}$ can be also understood from the chain of inequalities for different length scales in our problem: $l_\mathrm{B}\ll l_\mathrm{free}\ll l_\mathrm{CPI}\ll l_g$. Here $l_\mathrm{B}=p_{\perp}/eB\sim 3/\tilde{B}_0 T$ is the Larmor radius and $\tilde{B}_0=B_0/T_0^2$ is the dimensionless magnetic field. Below in Sec.~\ref{sec:NUMSOL}, we take $\tilde{B}_0= (10^{-1} - 10^{-2})$. We also use $l_\mathrm{CPI}\sim (\alpha_\mathrm{em}\mu_5)^{-1}$, which is the length scale of the chiral plasma instability~\cite{Akamatsu:2013pjd}. Finally, $l_g\sim \sigma_\mathrm{cond}(\alpha_\mathrm{em}\mu_5)^{-2}$ is the anomaly growth time scale~\cite{Pavlovich}. In strong magnetic fields, the first condition $l_\mathrm{B}\ll l_\mathrm{free}$ is always fulfilled, and obviously, $\mu_5\ll \alpha_\mathrm{em}T$ is the real condition in hot plasma resulting from the inequality $l_\mathrm{free}\ll l_\mathrm{CPI}$ here. We agree with clear arguments in Ref.~\cite{Pavlovich} that in the absence of CME---i.e. when $\mu_5=0$---the fluid turbulence exists already at the background level in standard MHD. Thus, $\tau_d$ should be the main scale parameter to zeroth-order approximation.

\section{Kinetic equations for the magnetic energy and helicity spectra\label{sec:KINETIC}}

Based on the master Eqs.~(\ref{Faraday}) and~(\ref{velocity}), we derive the kinetic equations for the spectra of the magnetic energy $\mathcal{E}_\mathrm{B}=\mathcal{E}_\mathrm{B}(k,t)$
and the density of the magnetic helicity $\mathcal{H}_\mathrm{B}=\mathcal{H}_\mathrm{B}(k,t)$ analogously as in Refs.~\cite{Dvornikov:2014uza,Semikoz:2016lqv}:
\begin{align}
  \frac{\partial\mathcal{E}_\mathrm{B}}{\partial t} & =
  -2k^{2}\eta_\mathrm{eff}\mathcal{E}_\mathrm{B}+\alpha_{+}k^{2}\mathcal{H}_\mathrm{B},
  \label{kineticsE}
  \\
  \frac{\partial\mathcal{H}_\mathrm{B}}{\partial t} & =
  -2k^{2}\eta_\mathrm{eff}\mathcal{H}_\mathrm{B}+4\alpha_{-}\mathcal{E}_\mathrm{B},
  \label{kineticsH}
\end{align}
where
\begin{align}\label{parameters}
  \eta_\mathrm{eff} = &
  \eta_m+\frac{4}{3}\frac{\tau_{d}}{P+ \rho}
  \int \mathrm{d}p\mathcal{E}_\mathrm{B},
  \quad
  \alpha_{\pm}  =
  \alpha_\mathrm{CME} \mp \alpha_{d},
  \nonumber
  \\
  \alpha_\mathrm{CME} = & \frac{\Pi(t)}{\sigma_\mathrm{cond}},
  \quad
  \alpha_{d} = \frac{2}{3}\frac{\tau_{d}}{P + \rho}
  \int \mathrm{d}p p^{2}\mathcal{H}_\mathrm{B},
\end{align}
and $\Pi(t)=2\alpha_\mathrm{em}\mu_{5}(t)/\pi$ is the CME parameter. Note that the anomalous current $\mathbf{J}_\mathrm{CME}$ does not contribute to the drag velocity $\mathbf{v}$ in Eq.~(\ref{velocity}). The details of the derivation of Eqs.~\eqref{kineticsE}-\eqref{parameters} are provided in Appendix~\ref{sec:KINDER}.

The difference of our results from the findings of Ref.~\cite{Boyarsky:2011uy} is seen from the second nonlinear terms in Eq. (\ref{parameters}), which contain the drag time $\tau_{d}\sim\alpha_\mathrm{em}^{-2}/T$ when we take into account the turbulent motion $\sim \mathbf{v}$. Note that the effective magnetic diffusion coefficient $\eta_\mathrm{eff}$ in Eq.~(\ref{parameters}) coincides with that in Ref.~\cite{Campanelli:2007tc} (accounting also for the factor $P+\rho$ in the denominator missed there). The analog of the $\alpha$-dynamo parameter $\alpha_{\pm}$ in Eq.~(\ref{parameters}) differs from that derived in Ref.~\cite{Campanelli:2007tc}, mainly because of the absence of CME term there, and due to different signs ($\pm$) in turbulent contributions for the evolution of spectra $\mathcal{E}_\mathrm{B}$ and $\mathcal{H}_\mathrm{B}$ instead of the same sign ($+$) in both equations. 

Integrating Eq.~\eqref{kineticsE} and~\eqref{kineticsH} over the spectrum, we get the following evolution
equations:
\begin{align}\label{eq:EBHB}
  \frac{\mathrm{d}E_{\mathrm{B}}}{\mathrm{d}t} = &  
  \alpha_{\mathrm{CME}}\int\mathrm{d}k\, k^{2}\mathcal{H}_{\mathrm{B}}(k,t) -
  2\eta_m\int\mathrm{d}k\, k^{2}\mathcal{E}_{\mathrm{B}}(k,t),
  \nonumber
  \\
  & -
  \frac{2\tau_{d}}{3(P+\rho)}\int\mathrm{d}k\mathrm{d}p\, k^{2}
  \left[
    4\mathcal{E}_{\mathrm{B}}(k,t)\mathcal{E}_{\mathrm{B}}(p,t) +
    p^{2}\mathcal{H}_{\mathrm{B}}(k,t)\mathcal{H}_{\mathrm{B}}(p,t)
  \right],
  \nonumber
  \displaybreak[1] 
  \\
  \frac{\mathrm{d}H_{\mathrm{B}}}{\mathrm{d}t} = & 
  4\alpha_{\mathrm{CME}}
  \int\mathrm{d}k\,\mathcal{E}_{\mathrm{B}}(k,t) -
  2\eta_m\int\mathrm{d}k\, k^{2}\mathcal{H}_{\mathrm{B}}(k,t),
\end{align}
where
\begin{equation}\label{eq:EBHBdef}
  E_{\mathrm{B}}(t) = \int\mathrm{d}k\,\mathcal{E}_{\mathrm{B}}(k,t),
  \quad
  H_{\mathrm{B}}(t)=\int\mathrm{d}k\,\mathcal{H}_{\mathrm{B}}(k,t),
\end{equation}
are the magnetic energy density and the helicity density. It is interesting
to note that the matter turbulence directly contributes only to the
evolution of the magnetic energy, whereas the dependence of the helicity density $H_{\mathrm{B}}(t)$ on $\tau_d$ is indirect, being hidden in the first term $E_{\mathrm{B}}(t)$ that is proportional to $\alpha_\mathrm{CME}$ for  the derivative $\dot{H}_{\mathrm{B}}$ in Eq.~(\ref{eq:EBHB}). One can see that the only source of the instability in
Eq.~\eqref{eq:EBHB} is the CME. If we set $\alpha_{\mathrm{CME}}=0$ in Eq.~(\ref{eq:EBHB}),
one can see that both $\dot{E}_{\mathrm{B}}$ and $\dot{H}_{\mathrm{B}}$
are negative, and hence only the dissipation of the magnetic field is present in the system provided by the finite electric conductivity $\eta_m\neq 0$ for both $E_{\mathrm{B}}(t)$ and $H_{\mathrm{B}}(t)$ and additionally by the fluid turbulence $\sim \tau_d$ for the magnetic energy density
$E_{\mathrm{B}}(t)$.

\subsection{Representation of kinetic equations in the form of integral equations\label{sec:ANALSOL}}

Supposing that the parameters $\eta_\mathrm{eff}$ and $\alpha_{\pm}$ are slowly varying functions, we can represent Eqs.~\eqref{kineticsE} and~\eqref{kineticsH} in an alternative form which is useful for the comparison with the results of Ref.~\cite{Campanelli:2007tc}. Let us choose the initial condition in the form:
$\mathcal{E}_\mathrm{B}(k,t_{0})=\mathcal{E}_{0}(k)$ and $\mathcal{H}_\mathrm{B}(k,t_{0})=2q\mathcal{E}_{0}(k)/k$, where $0 \leq q \leq 1$, and $\mathcal{E}_{0}(k)$ is the arbitrary function.
Then, if $|\alpha_\mathrm{CME}|>|\alpha_{d}|$, one has
\begin{align}\label{eq:coshsinh}
  \mathcal{E}_\mathrm{B}(k,t) = &
  \mathcal{E}_{0}(k)\exp
  \left(
    -2k^{2} l^2_\mathrm{diss} 
  \right)
  \nonumber
  \\
  & \times
  \left[
    \cosh
    \left(
      2k l_\mathrm{CME}
    \right) +
    q\sqrt{
    \frac{\alpha_\mathrm{CME}-\alpha_{d}}
    {\alpha_\mathrm{CME}+\alpha_{d}}}
    \sinh
    \left(
      2k l_\mathrm{CME}
    \right)
  \right],
  \nonumber
  \\
  \mathcal{H}_\mathrm{B}(k,t)= &
  \frac{2\mathcal{E}_{0}(k)}{k}\exp
  \left(
    -2k^{2} l^2_\mathrm{diss} 
  \right)
  \nonumber
  \\
  & \times
  \left[
    q\cosh
    \left(
      2k l_\mathrm{CME}
    \right) +
    \sqrt{\frac{\alpha_\mathrm{CME}+\alpha_{d}}
    {\alpha_\mathrm{CME}-\alpha_{d}}}
    \sinh
    \left(
      2k l_\mathrm{CME}
    \right)
  \right].
\end{align}
In the opposite case, when $|\alpha_\mathrm{CME}|<|\alpha_{d}|$, the following representation is valid:
\begin{align}\label{eq:cossin}
  \mathcal{E}_\mathrm{B}(k,t)= &
  \mathcal{E}_{0}(k)\exp
  \left(
    -2k^{2} l^2_\mathrm{diss} 
  \right)
  \nonumber
  \displaybreak[1]
  \\
  & \times
  \left[
    \cos
    \left(
      2k l_d
    \right) +
    q\sqrt{\frac{\alpha_{d}-\alpha_\mathrm{CME}}
    {\alpha_{d}+\alpha_\mathrm{CME}}}
    \sin
    \left(
      2k l_d
    \right)
  \right],
  \nonumber
  \displaybreak[1]
  \\
  \mathcal{H}_\mathrm{B}(k,t)= &
  \frac{2\mathcal{E}_{0}(k)}{k}
  \exp
  \left(
    -2k^{2} l^2_\mathrm{diss} 
  \right)
  \nonumber
  \\
  & \times
  \left[
    q\cos
    \left(
      2k l_d
    \right) -
    \sqrt{\frac{\alpha_{d}+\alpha_\mathrm{CME}}
    {\alpha_{d}-\alpha_\mathrm{CME}}}
    \sin
    \left(
      2k l_d
    \right)
  \right].
\end{align}
In the special situation of the aperiodic attenuation, if $|\alpha_\mathrm{CME}|=|\alpha_{d}|$, one can write down that
\begin{align}
  \mathcal{E}_\mathrm{B}(k,t) = &
  \mathcal{E}_{0}(k)
  \exp
  \left(
    -2k^{2} l^2_\mathrm{diss} 
  \right),
  \nonumber
  \displaybreak[1]
  \\
  \mathcal{H}_\mathrm{B}(k,t) = &
  \mathcal{E}_{0}(k)
  \exp
  \left(
    -2k^{2} l^2_\mathrm{diss} 
  \right)
  \left[
    8l^{(0)}_\mathrm{CME} +
    \frac{2q}{k}
  \right],
\end{align}
when $\alpha_\mathrm{CME}=\alpha_{d}$, and
\begin{align}\label{eq:exp-}
  \mathcal{E}_\mathrm{B}(k,t) = &
  \mathcal{E}_{0}(k)
  \exp
  \left(
    -2k^{2} l^2_\mathrm{diss} 
  \right)
  \left[
    4q l^{(0)}_\mathrm{CME}k +1
  \right],
  \nonumber
  \displaybreak[1]
  \\
  \mathcal{H}_\mathrm{B}(k,t) = &
  \frac{2q\mathcal{E}_{0}(k)}{k}
  \exp
  \left(
    -2k^{2} l^2_\mathrm{diss} 
  \right),
\end{align}
if $\alpha_\mathrm{CME}=-\alpha_{d}$. We use the following notations:
\begin{align}
  l_\mathrm{CME} = &
  \int_{t_{0}}^t \sqrt{\alpha_\mathrm{CME}^{2}(t')-\alpha_{d}^{2}(t')} \mathrm{d}t',
  \quad
  l^{(0)}_\mathrm{CME} =
  \int_{t_{0}}^t \alpha_\mathrm{CME}(t') \mathrm{d}t',
  \notag
  \\
  l_d = &
  \int_{t_{0}}^t \sqrt{\alpha_{d}^{2}(t') - \alpha_\mathrm{CME}^{2}(t')} \mathrm{d}t',
  \quad
  l^2_\mathrm{diss} =
  \int_{t_{0}}^t \eta_\mathrm{eff}(t') \mathrm{d}t',
  \quad
\end{align}
in Eqs.~\eqref{eq:coshsinh}-\eqref{eq:exp-}.

It should be noted that the solution of the kinetic Eqs.~\eqref{kineticsE} and~\eqref{kineticsH} considered in Ref.~\cite{Campanelli:2007tc} corresponds to the case when $\alpha_\mathrm{CME} = 0$. One can see in Eq.~\eqref{eq:cossin} that there is no amplification of the magnetic field in this situation. If $\alpha_\mathrm{CME} = 0$, the magnetic field is oscillatory, attenuated by the effective magnetic diffusion $\eta_\mathrm{eff}$. In general, the parameters $\alpha_{d}$, especially $\alpha_\mathrm{CME}\sim \mu_5(t)$, are changed over time. To take into account this fact, we should look for numerical solutions of the nonlinear kinetic Eqs.~(\ref{kineticsE}) and~\eqref{kineticsH}.

\subsection{Numerical solution to kinetic equations\label{sec:NUMSOL}}

When we study the evolution of magnetic fields in a hot plasma in the expanding Universe, it is convenient to rewrite Eqs.~\eqref{kineticsE}-\eqref{parameters} using the conformal dimensionless variables. They are introduced in the following way: $t\to \eta=M_0/T$ and $\tilde{k}=ak$, where $a=1/T$; $M_0=M_\mathrm{Pl}/1.66\sqrt{g^*}$; $M_\mathrm{Pl}=1.2\times 10^{19}\,\mathrm{GeV}$ is the Planck mass; and $g^{*}=106.75$ is the number of the relativistic degrees of
freedom. In these variables, Eqs.~\eqref{kineticsE} and~\eqref{kineticsH}
take the form
\begin{eqnarray}\label{eq:EBHBt}
  &&\frac{\partial\tilde{\mathcal{E}}_\mathrm{B}}{\partial\eta} =
  -2\tilde{k}^{2}\tilde{\eta}_\mathrm{eff}\tilde{\mathcal{E}}_\mathrm{B} +
  \tilde{\alpha}_{+}\tilde{k}^{2}\tilde{\mathcal{H}}_\mathrm{B},
  \nonumber
  \\
  &&\frac{\partial\tilde{\mathcal{H}}_\mathrm{B}}{\partial\eta} =
  - 2\tilde{k}^{2}\tilde{\eta}_\mathrm{eff}\tilde{\mathcal{H}}_\mathrm{B} +
  4\tilde{\alpha}_{-}\tilde{\mathcal{E}}_\mathrm{B}.
\end{eqnarray}
Here $\tilde{\mathcal{E}}_\mathrm{B}=\tilde{\mathcal{E}}_\mathrm{B}(\tilde{k},\eta)$ and
$\tilde{\mathcal{H}}_\mathrm{B}=\tilde{\mathcal{H}}_\mathrm{B}(\tilde{k},\eta)$ are the conformal spectra,
and
\begin{align}
  \tilde{\eta}_\mathrm{eff} = & \frac{\eta_\mathrm{eff}}{a} =
  \sigma_{c}^{-1} +
  \frac{4}{3}\frac{\alpha_\mathrm{em}^{-2}}{\tilde{\rho} +\tilde{p}}
  \int \mathrm{d}\tilde{p}\tilde{\mathcal{E}}_\mathrm{B},
  \nonumber
  \displaybreak[1]
  \\
  \tilde{\alpha}_{\pm} = & \alpha_{\pm} =
  \frac{\tilde{\Pi}}{\sigma_{c}} \mp
  \frac{2}{3}\frac{\alpha_\mathrm{em}^{-2}}{\tilde{\rho}+\tilde{p}}
  \int \mathrm{d}\tilde{p}\tilde{p}^{2}\tilde{\mathcal{H}}_\mathrm{B}.
\end{align}
The CME parameter takes the form
\begin{equation}
  \tilde{\Pi}=a\Pi=\frac{2\alpha_\mathrm{em}}{\pi}\tilde{\mu}_{5},
\end{equation}
and in a hot relativistic plasma one substitutes
\begin{equation}
  P=\frac{\rho}{3},\quad\rho=\frac{\pi^{2}}{30}g^{*}T^{4},
\end{equation}
or $P+\rho=2\pi^{2}g^{*}T^{4}/45$ and $\tilde{P}+\tilde{\rho}=2\pi^{2}g^{*}/45$.

The evolution equation for the chiral imbalance $\tilde{\mu}_{5}=\tilde{\mu}_{5}(\eta)$
has the form
\begin{equation}\label{mu5}
  \frac{\mathrm{d}\tilde{\mu}_{5}}{\mathrm{d}\eta} =
  -\frac{6\alpha_\mathrm{em}}{\pi}
  \int \mathrm{d}\tilde{k}
  \frac{\partial\tilde{\mathcal{H}}_\mathrm{B}}{\partial\eta} -
  \tilde{\Gamma}_{f}\tilde{\mu}_{5},
\end{equation}
where 
\begin{equation}
  \tilde{\Gamma}_{f}=a\Gamma_{f}=\alpha_\mathrm{em}^{2}
  \left(
    \frac{m_{e}}{3M_{0}}
  \right)^{2}
  \eta^{2}
\end{equation}
is the helicity flip rate~\cite{Boyarsky:2011uy}.

Before we analyze the general case numerically, let us discuss
the approximation of the monochromatic spectrum,
\begin{equation}
  \mathcal{E}_{\mathrm{B}}(\tilde{k},\eta) =
  \mathcal{\tilde{E}}_{0}(\eta)\delta(\tilde{k}-\tilde{k}_{0}),
  \quad
  \mathcal{H}_{\mathrm{B}}(\tilde{k},\eta) =
  \mathcal{\tilde{H}}_{0}(\eta)\delta(\tilde{k}-\tilde{k}_{0}),
\end{equation}
where $\tilde{k}_{0}$ is a characteristic conformal momentum, and $\mathcal{\tilde{E}}_{0}$ and $\mathcal{\tilde{H}}_{0}$ are new
unknown functions. The evolution equations~\eqref{eq:EBHBt} and~\eqref{mu5} take the form
\begin{align}\label{eq:EHmu5}
  \frac{\mathrm{d}\tilde{\mathcal{E}}_{0}}{\mathrm{d}\eta} = &
  -\frac{2\tilde{k}_{0}^{2}}{\sigma_{c}}\tilde{\mathcal{E}}_{0} +
  \frac{2\alpha_{\mathrm{em}}\tilde{k}_{0}^{2}}{\pi\sigma_{c}}
  \tilde{\mu}_{5}\tilde{\mathcal{H}}_{0} -
  \frac{2}{3}\tilde{\xi}\tilde{k}_{0}^{2}
  \left[
    4\mathcal{\tilde{E}}_{0}^{2}+\tilde{k}_{0}^{2}\mathcal{\tilde{H}}_{0}^{2}
  \right],
  \nonumber
  \\
  \frac{\mathrm{d}\tilde{\mathcal{H}}_{0}}{\mathrm{d}\eta} = &
  -\frac{2\tilde{k}_{0}^{2}}{\sigma_{c}}\tilde{\mathcal{H}}_{0} +
  \frac{8\alpha_{\mathrm{em}}}{\pi\sigma_{c}}
  \tilde{\mu}_{5}\tilde{\mathcal{E}}_{0},
  \nonumber
  \\
  \frac{\mathrm{d}\tilde{\mu}_{5}}{\mathrm{d}\eta} = &
  -\frac{6\alpha_{\mathrm{em}}}{\pi}
  \frac{\mathrm{d}\tilde{\mathcal{H}}_{0}}{\mathrm{d}\eta} -
  \tilde{\Gamma}_{f}\tilde{\mu}_{5},
\end{align}
where $\tilde{\xi}=(45/2g^{*})(\alpha_{\mathrm{em}}\pi)^{-2}$ is the turbulence parameter coming from the velocity field ${\bf v}\sim \tau_d$ in Eq.~(\ref{velocity}).

Using the new variables,
\begin{equation}\label{eq:newvarmon}
  \tau=\frac{2\tilde{k}_{0}^{2}}{\sigma_{c}}\eta,
  \quad
  R(\tau)=\frac{24\alpha_{\mathrm{em}}^{2}}{\pi^{2}\tilde{k}_{0}^{2}}
  \mathcal{\tilde{E}}_{0}(\eta),
  \quad
  H(\tau)=\frac{12\alpha_{\mathrm{em}}^{2}}{\tilde{k}_{0}\pi^{2}}
  \tilde{\mathcal{H}}_{0}(\eta),
  \quad
  M(\tau)=\frac{2\alpha_{\mathrm{em}}}{\pi\tilde{k}_{0}}\tilde{\mu}_{5}(\eta),
\end{equation}
Eq.~(\ref{eq:EHmu5}) can be rewritten as
\begin{align}\label{eq:RHM}
  \frac{\mathrm{d}R}{\mathrm{d}\tau}= & -R+MH-\xi(R^{2}+H^{2}),
  \nonumber
  \\
  \frac{\mathrm{d}H}{\mathrm{d}\tau}= & -H+MR,
  \nonumber
  \\
  \frac{\mathrm{d}M}{\mathrm{d}\tau}= & H-MR-GM,
\end{align}
where
\begin{equation}
  \xi=\frac{5\pi^{2}\tilde{k}_{0}^{2}\sigma_{c}}{2\alpha_{\mathrm{em}}^{4}g^{*}},
  \quad
  G=\frac{\sigma_{c}\tilde{\Gamma}_{f}}{2\tilde{k}_{0}^{2}}.
\end{equation}
Equation~(\ref{eq:RHM}) should be completed with the initial condition
$R_{0}=R(\tau_{0})$, $H_{0}=H(\tau_{0})=qR_{0}$, where $0\leq q\leq1$,
and $M_{0}=M(\tau_{0})$.

The system in Eq.~\eqref{eq:RHM} can be solved analytically if we neglect the evolution of the
chiral imbalance; i.e., when we set $M=0$. For $q>0$, the solution of Eq.~(\ref{eq:RHM})
has the form,
\begin{equation}\label{eq:RHsol}
  R(\tau)=H_{0}e^{-\tau}\cot
  \left[
    \xi H_{0}\left(1-e^{-\tau}\right)+\varphi_{0}
  \right],
  \quad H(\tau)=H_{0}e^{-\tau},
\end{equation}
where $\tan\varphi_{0}=q=H_{0}/R_{0}$. If $q=0$, then
\begin{equation}\label{eq:RH0sol}
  R(\tau)=\frac{R_{0}e^{-\tau}}{\xi R_{0}\left(1-e^{-\tau}\right)+1},
  \quad
  H(\tau)=0.
\end{equation}
In Eqs.~(\ref{eq:RHsol}) and~(\ref{eq:RH0sol}) we assume that
$\tau_{0}=0$.

\begin{figure}
  \centering
  \includegraphics[scale=.12]{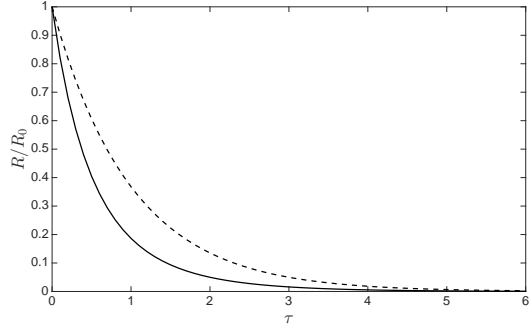}
  \protect
  \caption{\label{fig:monR}
  The normalized magnetic energy density $R/R_0$ versus $\tau$ on the basis of 
  Eq.~\eqref{eq:RHsol}. The solid line shows the evolution of $R$ accounting for the
  turbulence and corresponds to $\Delta_h = 0.5$. The dashed line corresponds to
  the situation without the turbulence.}
\end{figure}

To illustrate the behavior of the magnetic energy in Eq.~(\ref{eq:RHsol}),
in Fig.~\ref{fig:monR} we show $R(\tau)$ for $\Delta_{h}=\xi H_{0}=0.5$ versus $0<\tau<6$. In Fig.~\ref{fig:monR}, we suppose that $q=1$.
We also present the case when no turbulence is accounted for---i.e.,
when $\xi=0$, shown as the dashed line in Fig.~\ref{fig:monR}. One can see that the turbulent
motion of matter results in the faster decay of the magnetic field,
whereas the evolution of the magnetic helicity is not affected by
the turbulence; cf. Eq.~(\ref{eq:RHsol}). This result is in agreement
with our findings in Sec.~\ref{sec:KINETIC}, where the general case was studied. Indeed,
as one can see in Eq.~(\ref{eq:EBHB}), the contribution of the turbulence
terms to $\dot{E}_{\mathrm{B}}$ is negative; i.e., they cause $E_{\mathrm{B}}$
to decay faster than in the absence of the turbulence.

Now we turn to the study of the numerical solution of Eq.~\eqref{eq:EBHBt} in the general case. Let us use the initial energy spectrum in the form $\tilde{\mathcal{E}}_\mathrm{B}(\tilde{k},\eta_{0})=\mathcal{C}\tilde{k}^{\nu_\mathrm{B}}.$
The factor $\mathcal{C}$ can be found from the condition
\begin{equation}\label{initialenergy}
  \frac{\tilde{B}_{0}^{2}}{2} =
  \int \mathrm{d}\tilde{k}\tilde{\mathcal{E}}_\mathrm{B}(\tilde{k},\eta_{0}),
\end{equation}
where $\tilde{B}_{0}=\tilde{B}(\eta=\eta_{0})$ is the initial magnetic
field. If we use the Batchelor initial spectrum with $\nu_\mathrm{B}=4$ and $0<\tilde{k}<\tilde{k}_\mathrm{max}$, then, analogously to Eq.~\eqref{eq:newvarmon},
it is convenient to introduce the following dimensionless variables:
\begin{align}
  \mathcal{H}(\kappa,\tau) = &
  \frac{12\alpha_\mathrm{em}^{2}}{\pi^{2}}
  \tilde{\mathcal{H}}_\mathrm{B}(\tilde{k},\eta),
  \quad
  \mathcal{R}(\kappa,\tau) =
  \frac{24\alpha_\mathrm{em}^{2}}{\pi^{2}\tilde{k}_\mathrm{max}}
  \tilde{\mathcal{E}}_\mathrm{B}(\tilde{k},\eta),
  \quad
  \mathcal{M}(\tau) =
  \frac{2\alpha_\mathrm{em}}{\pi\tilde{k}_\mathrm{max}}\tilde{\mu}_{5}(\eta),
  \notag
  \\
  \tau = & \frac{2\tilde{k}_\mathrm{max}^{2}}{\sigma_{c}}\eta,
  \quad
  \kappa=\frac{\tilde{k}}{\tilde{k}_\mathrm{max}},
  \quad
  \mathcal{G} =
  \frac{\sigma_{c}}{2\tilde{k}_\mathrm{max}^{2}}\tilde{\Gamma}_{f}.
\end{align}
Using these variables, the system of kinetic equations takes the form,
\begin{align}
  \frac{\partial\mathcal{H}}{\partial\tau}=  &  
  -\kappa^{2}\mathcal{H}
  \left[
    1+K_{d}\int_{0}^{1}\mathrm{d}\kappa'\mathcal{R}(\kappa',\tau)
  \right]+
  \mathcal{R}
  \left[
    \mathcal{M}+
    K_{d}\int_{0}^{1}\mathrm{d}\kappa'\kappa^{\prime2}\mathcal{H}(\kappa',\tau)
  \right],
  \label{finalH}
  \\
  \frac{\partial\mathcal{R}}{\partial\tau}=  & 
  -\kappa^2\mathcal{R}
  \left[
    1+K_{d}\int_{0}^{1}\mathrm{d}\kappa'\mathcal{R}(\kappa',\tau)
  \right]+
  \kappa^{2}\mathcal{H}
  \left[
    \mathcal{M}-
    K_{d}\int_{0}^{1}\mathrm{d}\kappa'\kappa^{\prime2}\mathcal{H}(\kappa',\tau)
  \right],
  \label{finalE}
  \\
  \frac{\mathrm{d}\mathcal{M}}{\mathrm{d}\tau}=  &
  \int_{0}^{1}\mathrm{d}\kappa
  \left(
    \kappa^{2}\mathcal{H}-\mathcal{R}\mathcal{M}
  \right) -
  \mathcal{G}\mathcal{M},
  \label{finalM}
\end{align}
where $K_{d}=5\sigma_{c}\tilde{k}_\mathrm{max}^{2}/4\alpha_\mathrm{em}^{4}g^{*}$. It is interesting to note that the contribution of the turbulent terms cancels out in Eq.~\eqref{finalM}. Nevertheless there is a turbulence contribution in Eq.~\eqref{finalH} contrary to Eq.~\eqref{eq:RHM} valid for the monochromatic spectrum.

The initial values of the functions $\mathcal{R}$ and $\mathcal{H}$
are $\mathcal{R}(\kappa,\tau_{0})=\mathcal{R}_{0}\kappa^{\nu_\mathrm{B}}$
and $\mathcal{H}(\kappa,\tau_{0})=q\mathcal{R}_{0}\kappa^{\nu_\mathrm{B}-1}$,
where
\begin{equation}
  \mathcal{R}_{0} = \frac{12\alpha_\mathrm{em}^{2}\tilde{B}_{0}^{2}}
  {\pi^{2}\tilde{k}_\mathrm{max}^{2}}
  (\nu_\mathrm{B}+1).
\end{equation}
and correspondingly to the MHD bound on the magnetic helicity value \cite{Biscamp}, $0\leq q\leq1$.

\begin{figure}
  \centering
  \subfigure[]
  {\label{1a}
  \includegraphics[scale=.12]{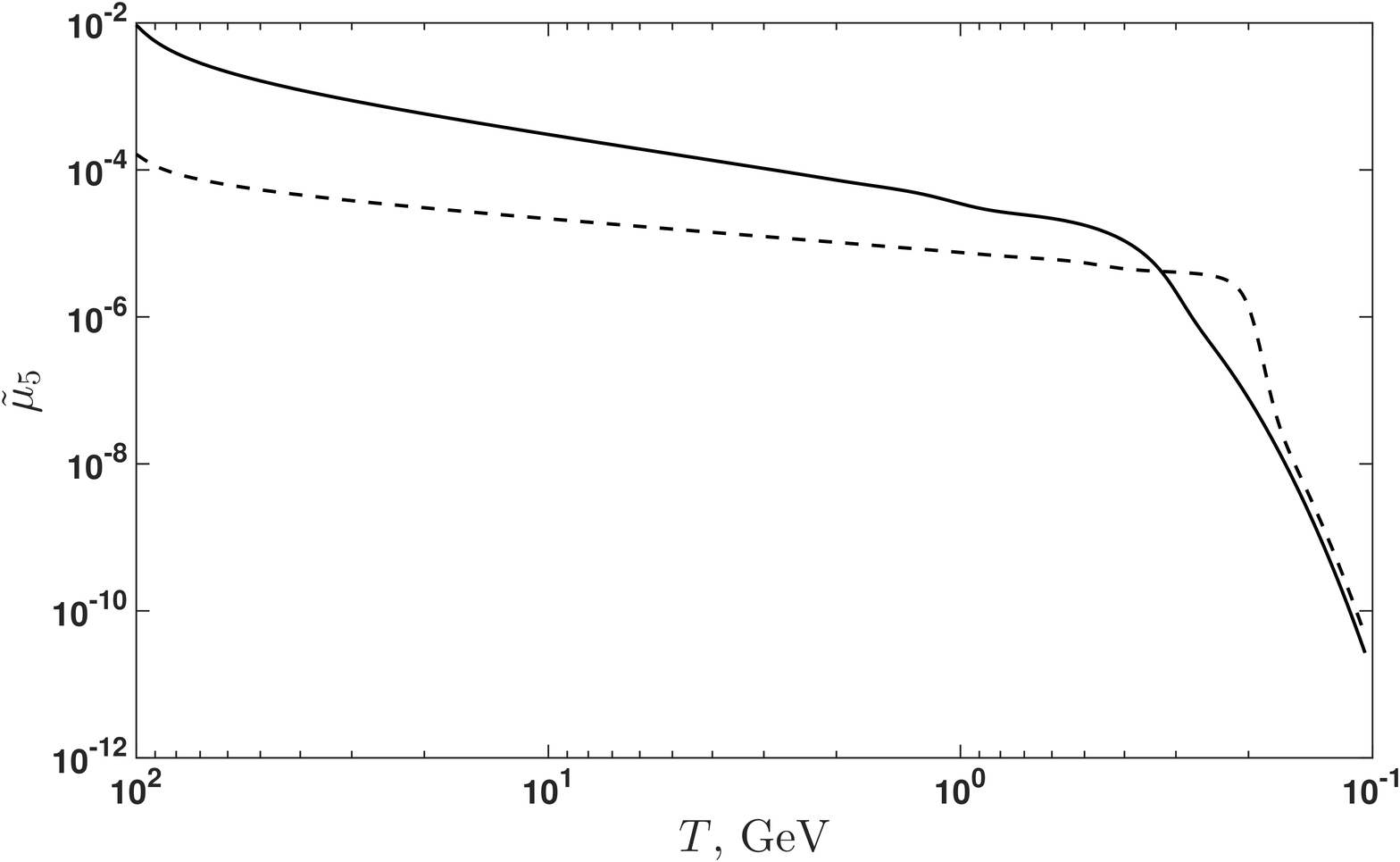}}
  \hskip-.7cm
  \subfigure[]
  {\label{1b}
  \includegraphics[scale=.12]{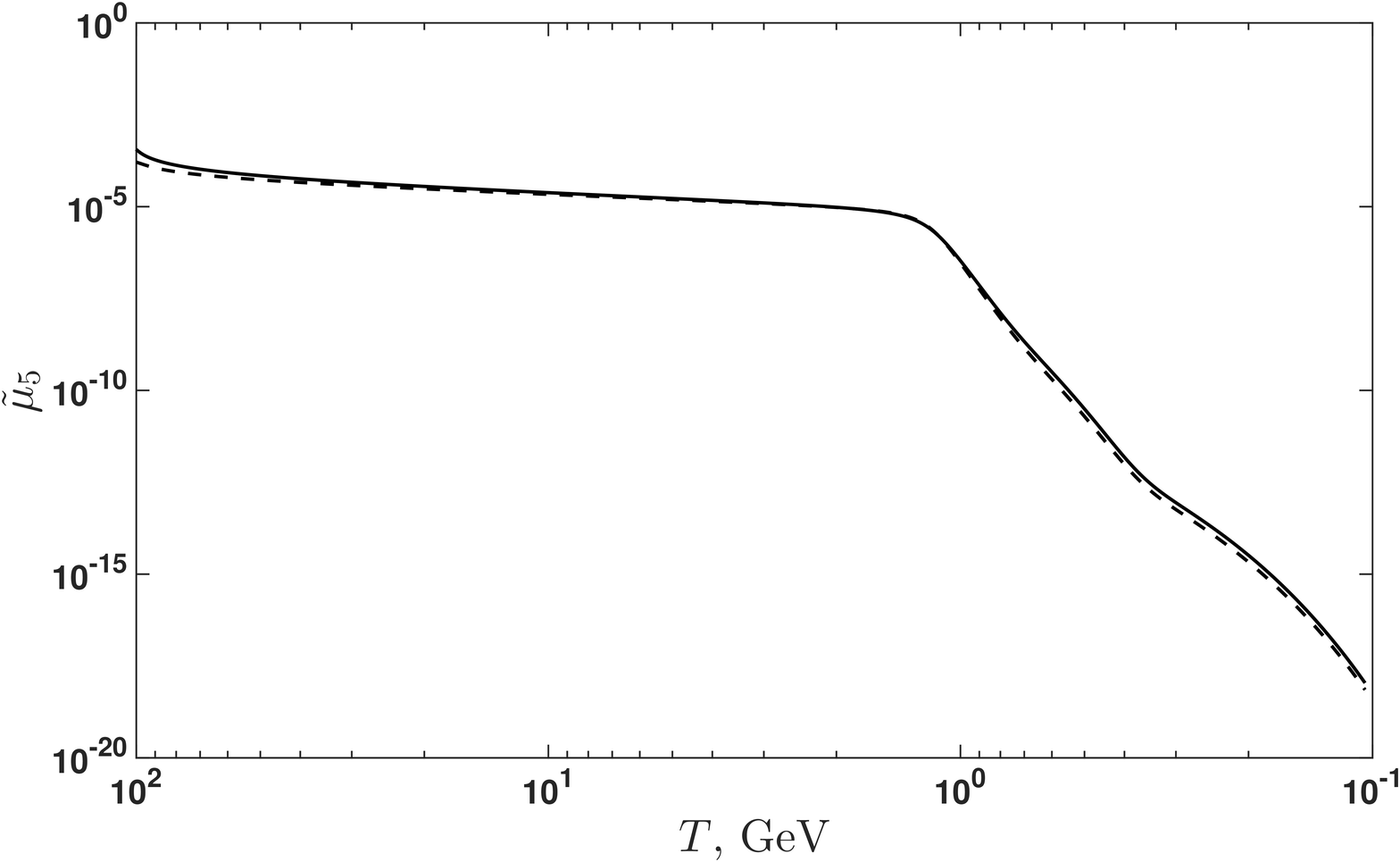}}
  \\
  \subfigure[]
  {\label{1c}
  \includegraphics[scale=.12]{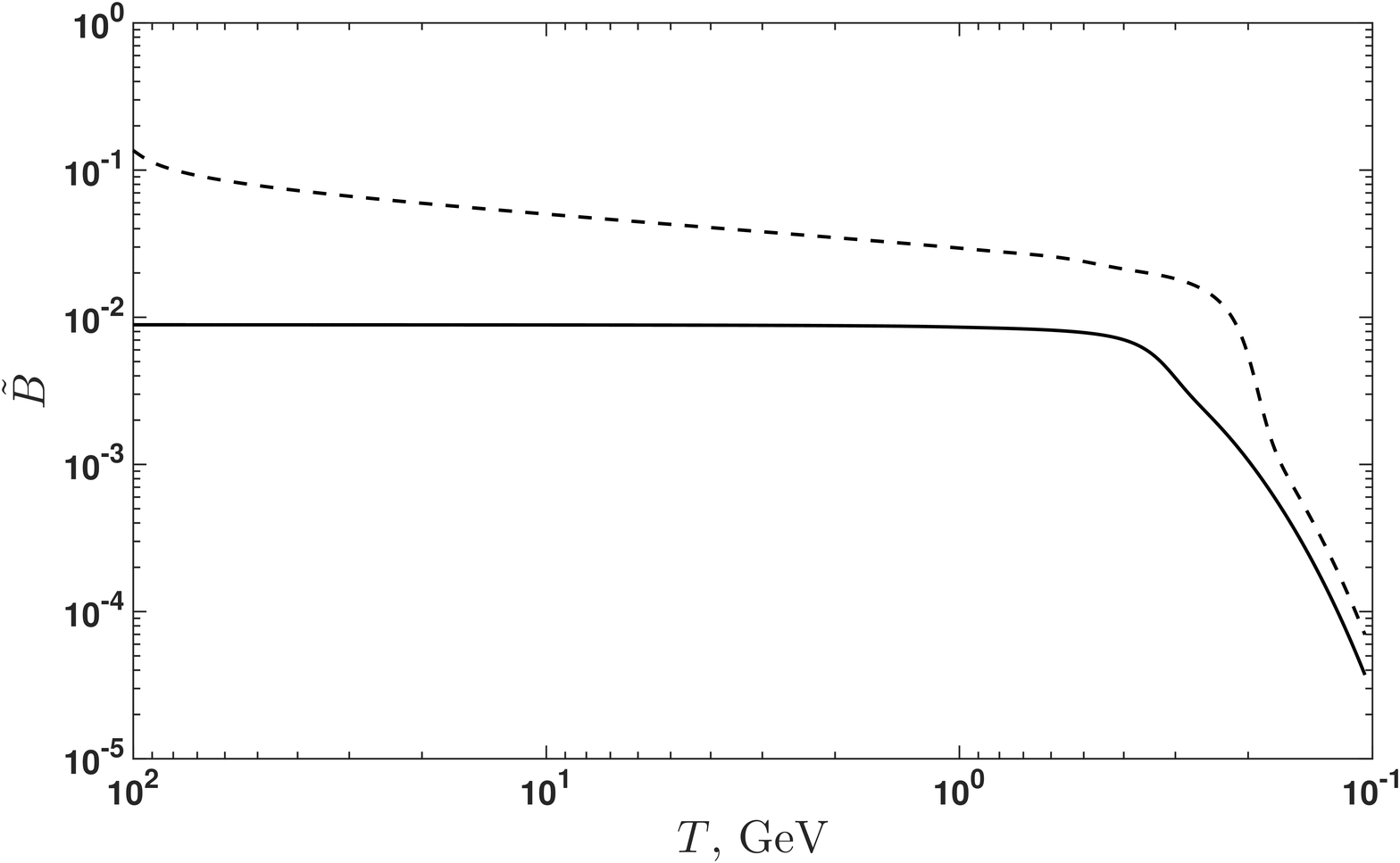}}
  \hskip-.7cm
  \subfigure[]
  {\label{1d}
  \includegraphics[scale=.12]{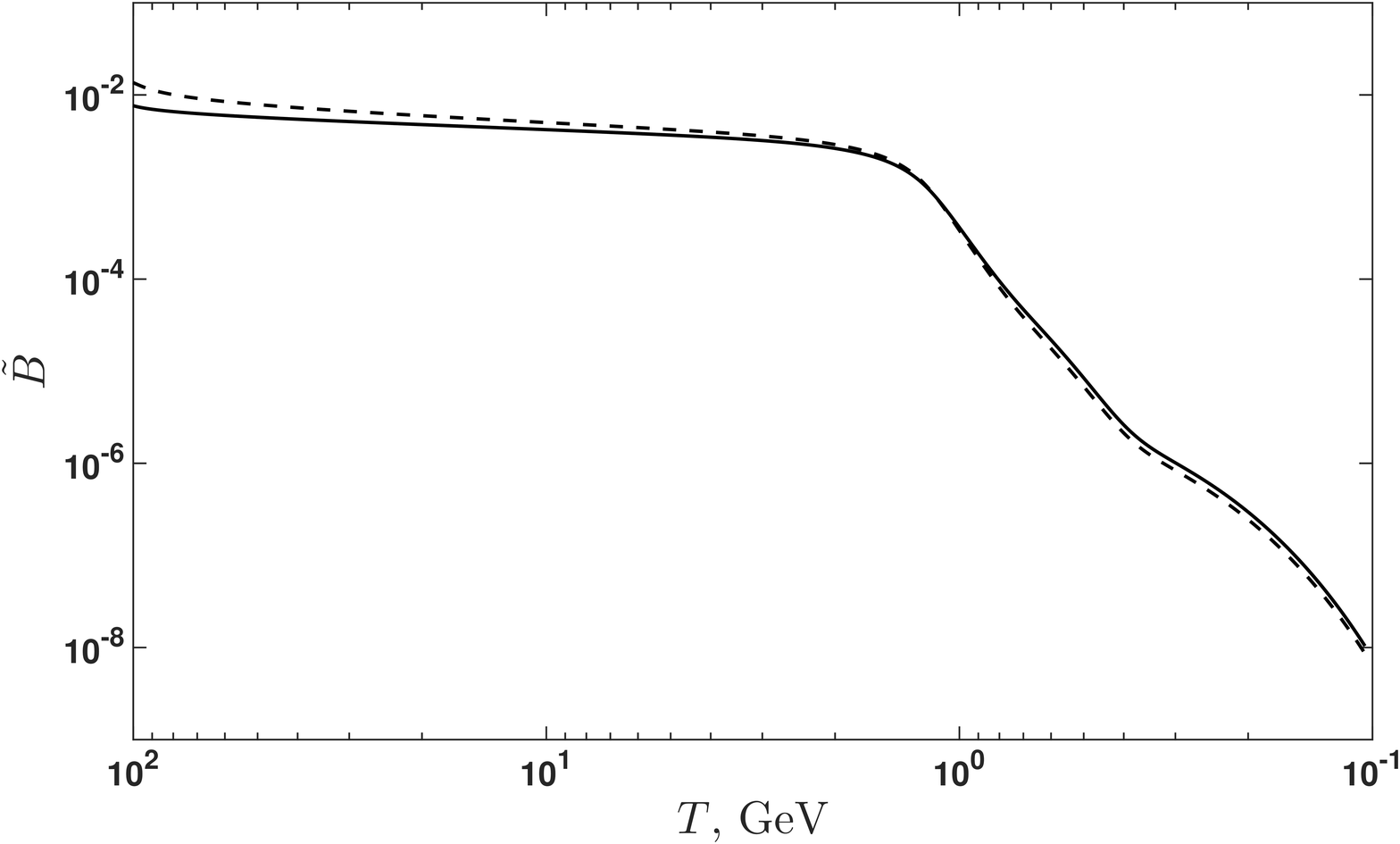}}
  \\
  \subfigure[]
  {\label{1e}
  \includegraphics[scale=.12]{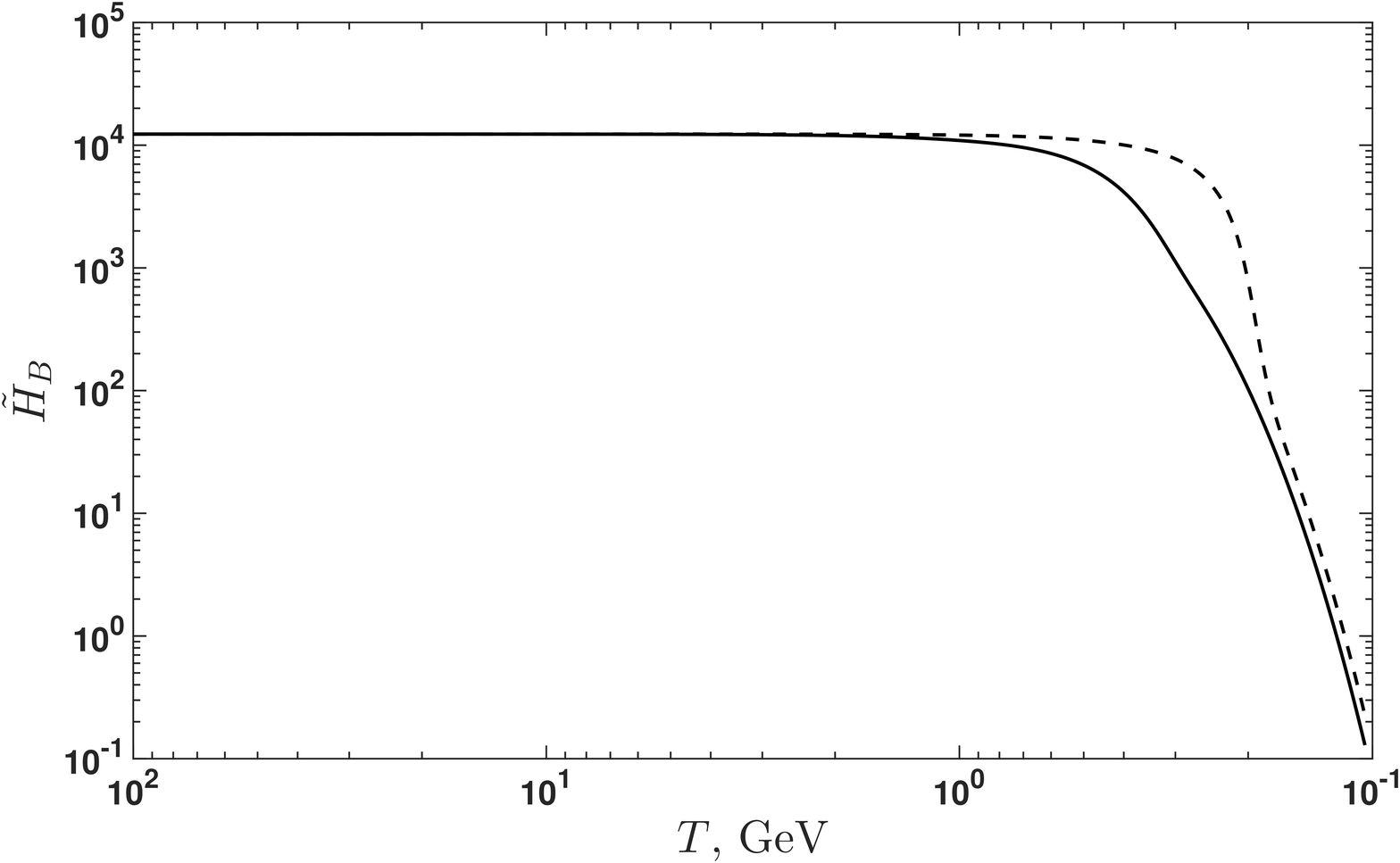}}
  \hskip-.7cm
  \subfigure[]
  {\label{1f}
  \includegraphics[scale=.12]{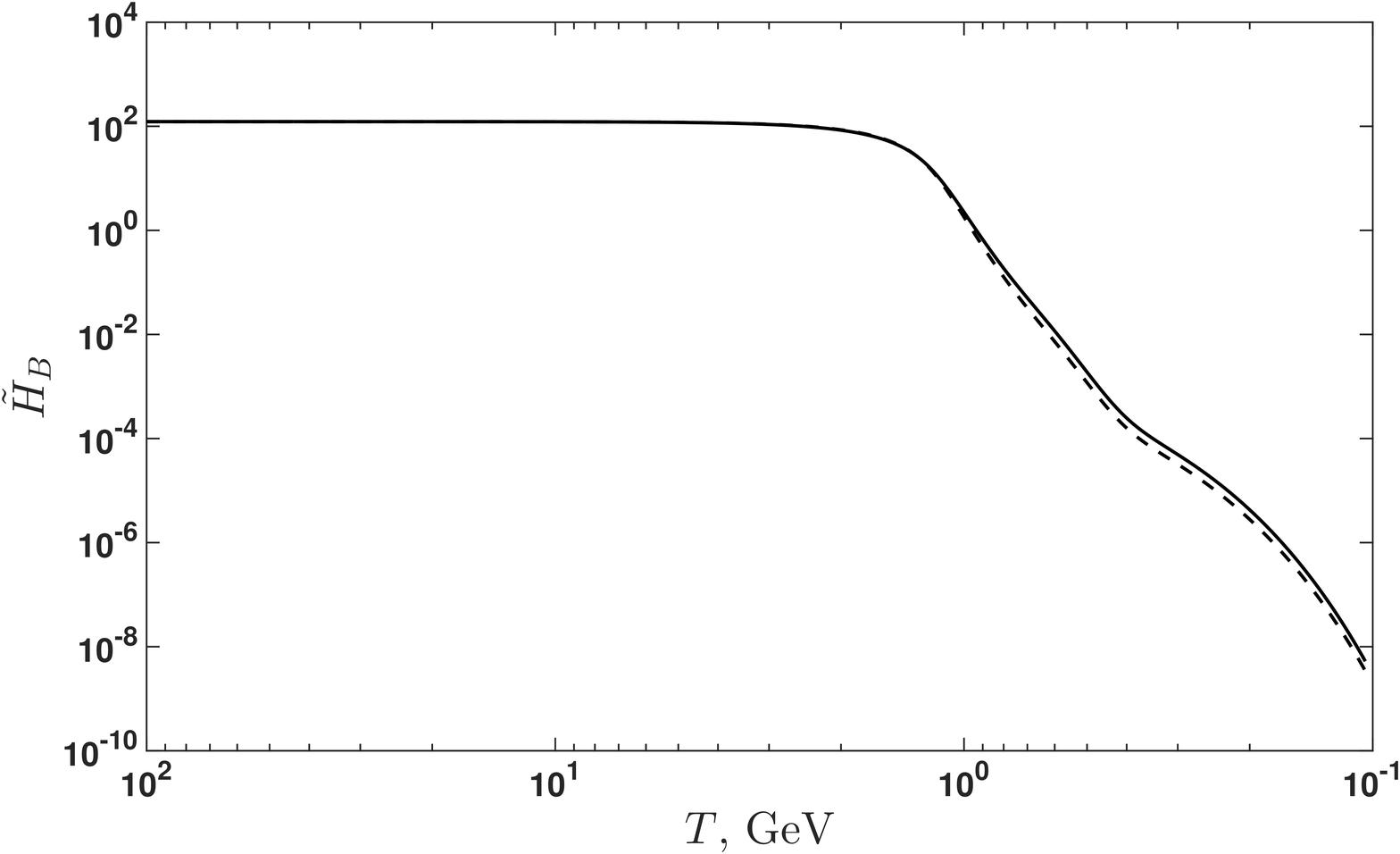}}
  \protect
  \caption{\label{fig:Bevol}
  The evolution of the chiral imbalance, the magnetic energy density
  and the helicity density in the plasma of the early Universe
  at $10^2\,\mathrm{MeV} < T < 10^2\,\mathrm{GeV}$.
  (a) and (b): The evolution of the chiral imbalance,
  $\mu_5(T)=(\mu_\mathrm{R} - \mu_\mathrm{L})/2$.
  (c) and (d):  The evolution of the magnetic field,
  $B=\sqrt{2\int \mathrm{d}k\mathcal{E}_\mathrm{B}(k,t)}$.
  (e) and (f):  The evolution  of the magnetic helicity density.
  Panels (a), (c), and (e) correspond to $\tilde{B}_0=10^{-1}$, whereas
  panels (b), (d), and (f) correspond to $\tilde{B}_0=10^{-2}$.
  Solid lines show the evolution accounting for both
  the turbulence effects ($K_d \neq 0$) and CME, whereas dashed lines show the  
  evolution for the CME case only ($K_d = 0$).}
\end{figure}

We solve kinetic equations~(\ref{finalH})--\eqref{finalM} numerically.
The influence of the turbulent matter motion $\sim \mathbf{v}$ on the MHD characteristics, such as the magnetic field strength and the magnetic helicity, as well as on the chiral asymmetry parameter $\mu_5(t)$ in a hot plasma in the broken phase of the early Universe is illustrated in Fig.~\ref{fig:Bevol}. The solid lines correspond to the case where both effects---i.e., CME and the turbulent motion of matter $\mathbf{v}\sim \tau_d$---are taken into account, while the dashed lines correspond to the CME effect only applied in Ref.~\cite{Boyarsky:2011uy}. Note that we present the numerical solutions of Eqs.~(\ref{finalH})--\eqref{finalM} for the maximum helicity parameter $q=1$ only. This means that we use the relation $\mathcal{H}_\mathrm{B}(k,t_0)=2\mathcal{E}_\mathrm{B}(k,t_0)/k$ for the initial Batchelor spectrum in Eq. (\ref{initialenergy}), where $0\leq \tilde{k}\leq \tilde{k}_\mathrm{max}=10^{-6}$. In Figs.~\ref{1a}, \ref{1c} and~\ref{1e}, we show results for the maximum initial magnetic field $\tilde{B}_0=0.1$ still obeying the BBN bound on the magnetic field $B\leq 10^{11}\,\mathrm{G}$ \cite{Cheng:1993kz} at the temperature $T_{{\rm BBN}}=0.1\,\mathrm{MeV}$~\cite{fn2}. In Figs.~\ref{1c}, \ref{1d} and~\ref{1f} we show the results for a smaller seed field $\tilde{B}_0=10^{-2}$.

It should be noted that Eqs.~(\ref{finalH})-\eqref{finalM} turn out to be stiff for the chosen parameters. The technique used to obtain the numerical solution of this system involves a multipoint implicit finite difference method. In this method, not all initial conditions result in a smooth behavior of the solution. Thus, we have to omit some initial part of the curves which reveal a nonsmooth behavior. That is why
Figs.~\ref{1a}--\ref{1d} look as if the initial conditions were different for the variables corresponding to the solid and dashed lines. Surprisingly, this inconsistency does not affect the evolution of the magnetic helicity density shown in Figs.~\ref{1e} and~\ref{1f}.

\section{Discussion\label{sec:DISC}}

One can see in Fig.~\ref{fig:Bevol} that the stronger the initial magnetic field, the more noticeable the difference between the turbulent and nonturbulent cases is. The chiral anomaly parameter $\mu_5(t)$  is supported by a matter turbulence at a higher level just starting from the EWPT time, then somewhere at a few hundred ${\rm MeV}$ reduces more smoothly and drops down a bit earlier than accounting for the CME effect only. This can be explained by the inverse cascade with an increase of large-scale contributions in spectra when the role of turbulent motions ceases. 

The dependence of $\mu_5$ on the turbulence parameter $K_d$ is not trivial, being hidden in Eq.~(\ref{finalM}). Such a hidden dependence comes rather from the magnetic field characteristics $\mathcal{H}$ and $\mathcal{R}$, which evidently depend on that parameter as seen in Eqs.~(\ref{finalH}) and~\eqref{finalE}. While the diffusion terms for the magnetic helicity $\mathcal{H}$ and the magnetic energy density spectra $\mathcal{R}$ are both enhanced  by turbulent motions $\sim K_d$, the instability (generation) terms $\sim \mathcal{M}\sim \mu_5$ are supplemented differently through the same parameter $K_d$. The magnetic helicity $\mathcal{H}$ is supported by turbulent motions even for a decreasing chiral anomaly $\mu_5$; cf. Figs.~\ref{1e} and~\ref{1f}. The magnetic energy $\mathcal{R}$ reduces additionally through the turbulent parameter $K_d$. This is a reason why the solid curve for magnetic field strength in Figs.~\ref{1c} and~\ref{1d} occurs below the dash curves corresponding to the pure CME effect.

Let us stress that such opposite contributions of the turbulent motion $\sim \mathbf{v}$ to the evolution of $\mathcal{H}$ and $\mathcal{R}$ come directly from different signs of the parameter $\alpha_d$ in Eq.~(\ref{parameters}), as we found in contrast to the results in Ref.~\cite{Campanelli:2007tc}. Another important result obtained in the present work is the examination of the possibility for the plasma turbulence to drive the magnetic field instability. In our work, we have approximated the plasma velocity by the Lorentz force; cf. Eq.~\eqref{velocity}. In frames of this model, using the results of Sec.~\ref{sec:ANALSOL}, we can see that, if one accounts for only the plasma turbulence contribution---i.e., assuming that $\alpha_d \neq 0$ and $\alpha_\mathrm{CME} = 0$---the initial magnetic field cannot be amplified. This result follows from Eq.~\eqref{eq:cossin}. This, our new finding confronts the statement of Ref.~\cite{Campanelli:2007tc}, where it is claimed that plasma turbulence described within the chosen model can provide the enhancement of a seed magnetic field.

The physical reason for the aforementioned discrepancy of our results with the findings of Ref.~\cite{Campanelli:2007tc} is based on the following fact: The model to account for the plasma velocity in the Faraday equation~\eqref{Faraday} implies the replacement $\mathbf{v} \to \mathbf{F}_\mathrm{L}$ in Eq.~\eqref{velocity}. The Lorentz force is known not to be able to linearly accelerate charged particles in plasma. Thus, self-sustained electric currents, which could generate an unstable magnetic field, cannot be excited in such plasma. This means that the instability of the magnetic field cannot be implemented if we choose this model to take into account the turbulent motion of matter, contrary to the claim in Ref.~\cite{Campanelli:2007tc}. Therefore, the representation of the spectra in terms of hyperbolic functions in Eq.~\eqref{eq:coshsinh} is possible only if CME is accounted for and its contribution is dominant; i.e., when $|\alpha_\mathrm{CME}| > |\alpha_d|$. The turbulence alone can provide only a faster decay of large-$k$ modes in the spectra; cf. Eqs.~\eqref{eq:EBHB} and~\eqref{eq:RHsol}. 

One can expect that the inclusion of the velocity field could influence the evolution of the right and left circularly polarized modes $B_{\pm}(k)$ coming from the Faraday equation~(\ref{Faraday}), cf. Ref.~\cite{Vachaspati:2016xji}, where such a velocity was not taken into account. There remains also an interesting possibility to replace the vanishing CME by the contribution of the axion field to MHD, as pointed out recently in Ref.~\cite{Long:2015cza} [see Eq.~(40b) there]. 

\section*{Acknowledgments}

We are thankful to D.~D.~Sokoloff for useful discussions, as well as to N.~Leite and G.~Sigl for emphasizing the velocity field in the advection term of the Faraday equation, which was a primary motivation  for our work. One of the authors (M.~D.) is grateful to the Tomsk State University Competitiveness Improvement Program and RFBR (Research Project No.~15-02-00293) for partial support. 

\appendix

\section{Turbulence contribution to the kinetic equations for the spectra\label{sec:KINDER}}

In this Appendix, we derive the kinetic equations for $\mathcal{E}_\mathrm{B}$ and $\mathcal{H}_\mathrm{B}$ used in Sec.~\ref{sec:KINETIC} and show their difference from analogous equations in Refs.~\cite{Campanelli:2007tc,Pavlovich}.

We shall start with the derivation of the equation for $\mathcal{H}_\mathrm{B}$. Let us neglect the contribution of CME to the magnetic field evolution. Then the Faraday equation~\eqref{Faraday} takes the form,
\begin{equation}
  \partial_t\mathbf{B} =
  \nabla\times(\mathbf{v}\times \mathbf{B}) + \eta_m\nabla^2\mathbf{B}.
\end{equation}
Using the Fourier representation for the velocity $\mathbf{v}=\tau_d(\mathbf{J}\times \mathbf{B})/(P + \rho)$ in Eq.~\eqref{velocity} we find the evolution equation for the magnetic field,
\begin{multline}\label{Faraday2}
  \partial_t B_i(\mathbf{k},t) + \eta_m k^2B_i(\mathbf{k},t) = 
  \varepsilon_{ijk}k_j \frac{\tau_d}{P+\rho }
  \int \frac{\mathrm{d}^3p}{(2\pi)^3}
  \int \frac{\mathrm{d}^3q}{(2\pi)^3}q_rB_s(\mathbf{q})
  \\
  \times
  \left[
    \varepsilon_{krs}B_n(\mathbf{p} - \mathbf{q})B_n(\mathbf{k} - \mathbf{p})  
    -\varepsilon_{rsm}B_k(\mathbf{k} - \mathbf{p})B_m(\mathbf{p} - \mathbf{q})
  \right].
\end{multline}
which coincides with the analogous result in Ref.~\cite{Campanelli:2007tc}, except for the factor $(P + \rho)^{-1}$ missed in Ref.~\cite{Campanelli:2007tc}.

Using the evolution equation for the vector potential
$\partial_t{\bf A}=(\mathbf{v}\times \mathbf{B}) - \eta_m\mathbf{J}$, where $\mathbf{J}=(\nabla\times \mathbf{B})$ is the electric current in MHD, one finds in the Fourier representation
\begin{align}\label{potential}
  \partial_tA_i(-\mathbf{k}, t) + \eta_m k^2A_i(-\mathbf{k},t) = &
  \varepsilon_{ikl}\varepsilon_{kmt}\varepsilon_{mst} \frac{\tau_d}{P + \rho}
  \int \frac{\mathrm{d}^3p}{(2\pi)^3}\int \frac{\mathrm{d}^3q}{(2\pi)^3} (-\mathrm{i}q_s)
  \nonumber
  \\
  &
  \times B_t(\mathbf{q},t)B_n(\mathbf{p},t)B_l(-\mathbf{k} -\mathbf{p} -\mathbf{q}).
\end{align}
In Eq.~\eqref{potential} we change the sign of the momentum $\mathbf{k}\to - \mathbf{k}$ in the argument of $A_i$~\cite{fn3}, meaning to apply the two-point correlator
\begin{equation}\label{correlator}
  \langle
    B_i(\mathbf{k},t)B_j(\mathbf{p},t)
  \rangle =
  \frac{(2\pi)^3}{2}\delta^{(3)}(\mathbf{k} + \mathbf{p})
  \left[
    (\delta_{ij} - \hat{k}_i\hat{k}_j)S(k,t) +
    \mathrm{i}\varepsilon_{ijk}\hat{k}_k A(k,t)
  \right],
\end{equation}
for the Faraday equation~(\ref{Faraday2}) multiplied by the potential $A_i(-\mathbf{k},t)$, then summed with Eq.~(\ref{potential}) multiplied by the magnetic field $B_i(\mathbf{k},t)$. In Eq.~\eqref{correlator}, the form factors $S(k,t)$ and $A(k,t)$ are related to the spectra
\begin{equation}\label{eq:EHSA}
  \mathcal{E}_\mathrm{B}(k,t)=k^2\frac{S(k,t)}{(2\pi)^2},
  \quad
  \mathcal{H}_\mathrm{B}(k,t)=k\frac{A(k,t)}{2\pi^2},
\end{equation}
obeying the kinetic equations~(\ref{kineticsE}) and~\eqref{kineticsH}. 

Using the Maxwell equation valid for any choice of the Fourier representation  $k^2A_i(\mathbf{k},t)=J_i(\mathbf{k},t)$ and neglecting the derivative $\partial^2_{t}A_i=0$ as usual in MHD, as well as choosing the Fourier representation as in Ref.~\cite{Campanelli:2007tc}, 
\begin{equation}
  B_j({\bf x},t)= \int \frac{\mathrm{d}^3q}{(2\pi)^3}
  e^{-\mathrm{i}\mathbf{q}{\bf x}} B_j(\mathbf{q},t),
\end{equation}
one obtains for the averaged sum of binary products
\begin{equation}
  \langle B_i(\mathbf{k})[\partial_t A_i(-\mathbf{k}) +
  \eta_m k^2]A_i(-\mathbf{k}) \rangle +
  \langle A_i(-\mathbf{k})[\partial_t B_i(\mathbf{k}) +
  \eta_m k^2B_i(\mathbf{k})] \rangle
\end{equation}
the evolution equation,
\begin{equation}\label{helicity-spectrum}
  \frac{(2\pi)^5\delta^{(3)}(0)}{2k^2}
  \left[
    \partial_t  + 2\eta_m k^2
  \right]
  \mathcal{H}_\mathrm{B}(k,t)= I_1 + I_2,
\end{equation}
where we use Eq.~\eqref{eq:EHSA}. The integrals $I_{1,2}$ in Eq.~\eqref{helicity-spectrum}, which read
\begin{align}
  I_1 = & 
  \varepsilon_{ikl}\varepsilon_{kmn}\varepsilon_{mst} \frac{\tau_d}{P + \rho}
  \int \frac{\mathrm{d}^3p}{(2\pi)^3}\int \frac{\mathrm{d}^3q}{(2\pi)^3}(-\mathrm{i}q_s)
  \nonumber
  \\
  & \times
  \langle B_t(\mathbf{q})B_n(\mathbf{p})B_l(-\mathbf{k} -\mathbf{p} -\mathbf{q})B_i(\mathbf{k})\rangle
  \nonumber
  \\
  & =
  \frac{(2\pi)^5\delta^{(3)}(0)}{2k^2}
  \left[
    -\frac{2\tau_d\dot{H}_\mathrm{B}(t)}{3(P + \rho)\eta_m}\mathcal{E}_\mathrm{B}(k,t) -
    \frac{4}{3}\frac{\tau_d}{(P + \rho)}E_\mathrm{B}(t)k^2\mathcal{H}_\mathrm{B}(k,t)
  \right],
  \label{I_1}
  \\
  I_2 = & \varepsilon_{ikl}\varepsilon_{iqg}k_jk_q \frac{\tau_d}{P + \rho}
  \int \frac{\mathrm{d}^3p}{(2\pi)^3}\int \frac{\mathrm{d}^3q}{(2\pi)^3}(\mathrm{i}q_r)
  \nonumber
  \\
  & \times
  \langle B_s(\mathbf{q})[\varepsilon_{krs}B_n(\mathbf{p} - \mathbf{q})
  B_n(\mathbf{k} -\mathbf{p})B_g(-\mathbf{k}) -
  \varepsilon_{rsm}B_k(\mathbf{k} -\mathbf{p})
  B_m(\mathbf{p} - \mathbf{q})B_g(- \mathbf{k})
  \rangle
  \nonumber
  \\
  & =
  \frac{(2\pi)^5\delta^{(3)}(0)}{2k^2}
  \left[
    -\frac{2\tau_d\dot{H}_\mathrm{B}(t)}{3(P + \rho)\eta_m}\mathcal{E}_\mathrm{B}(k,t) - 
    \frac{4}{3}\frac{\tau_d}{(P + \rho)}E_\mathrm{B}(t)k^2\mathcal{H}_\mathrm{B}(k,t)
  \right],
  \label{I_2}
\end{align}
result from the multiplication of Eq.~(\ref{potential}) by $B_i(\mathbf{k})$ and Eq.~(\ref{Faraday2}) by $A_i(-\mathbf{k}) = \mathrm{i} \varepsilon_{iqg} k_q B_g(-\mathbf{k}) / k^2$, correspondingly, when using the four-point correlator
\begin{multline}\label{four-point}
  \langle
    B_i(\mathbf{k})B_j(\mathbf{p})B_k(\mathbf{q})B_l(\mathbf{s})
  \rangle
  = 
  \langle
    B_i(\mathbf{k})B_j(\mathbf{p})
  \rangle
  \langle
    B_k(\mathbf{q})B_l(\mathbf{s})
  \rangle
  \\
  +
  \langle
    B_i(\mathbf{k})B_k(\mathbf{q})
  \rangle
  \langle
    B_j(\mathbf{p})B_l(\mathbf{s})
  \rangle
  +
  \langle
    B_i(\mathbf{k})B_l(\mathbf{s})
  \rangle
  \langle
    B_j(\mathbf{p})B_k(\mathbf{q})
  \rangle,
\end{multline}
in the same form as in Refs.~\cite{Campanelli:2007tc,Biscamp}. In Eqs.~\eqref{I_1} and~\eqref{I_2}, ${E}_\mathrm{B}(t)$ is the magnetic energy density defined in Eq.~\eqref{eq:EBHBdef}.

It is interesting to note that $I_1=I_2$, giving finally from Eq.~(\ref{helicity-spectrum})
\begin{equation}\label{helicity-spectrum2}
  \left[
    \partial_t  + 2\eta_m k^2
  \right]
  \mathcal{H}_\mathrm{B}(k,t) =
  -\frac{4}{3}\frac{\tau_d}{\eta_m(P+ \rho)}
  \dot{H}_\mathrm{B}(t)\mathcal{E}_\mathrm{B}(k,t) -
  \frac{8}{3}\frac{\tau_d}{P+ \rho} k^2 E_\mathrm{B}(t)
  \mathcal{H}_\mathrm{B}(k,t).
\end{equation}
Adding the CME term to Eq.~\eqref{helicity-spectrum2}
and accounting for the standard MHD relation $\dot{H}_\mathrm{B}(t)= - 2 \eta_m \smallint_0^{\infty} p^2\mathrm{d}p \mathcal{H}_\mathrm{B}(p,t)$, one gets Eq.~\eqref{kineticsH},
\begin{equation}\label{helicity-spectrum3}
  \frac{\partial \mathcal{H}_\mathrm{B}(k,t)}{\partial t} =
  - 2k^2\eta_\mathrm{eff}\mathcal{H}_\mathrm{B}(k,t)  +
  4\alpha_- \mathcal{E}_\mathrm{B}(k,t), 
\end{equation}
where $\alpha_- = \alpha_\mathrm{CME} + \alpha_d$, $\alpha_d(t)=2\tau_d\int_0^{\infty}\mathrm{d}pp^2\mathcal{H}_\mathrm{B}(p,t)/3(P+\rho)$, correspondingly to notations in Eq.~\eqref{parameters}. Let us stress the coincidence of signs of the turbulent term $\alpha_d(t)$ and the analogous $\alpha_\mathrm{B}(t)$ in Refs.~\cite{Campanelli:2007tc,Pavlovich}, $\alpha_d(t)=\alpha_\mathrm{B}(t)= -\tau_d\dot{{H}}_\mathrm{B}(t)/[3\eta_m(P+\rho)]$, resulting in the coincidence of our Eq. (\ref{helicity-spectrum3}) in the case $\mu_5=0$ and, e.g.,  Eq.~(8) in Ref.~\cite{Campanelli:2007tc}.

Equation~\eqref{kineticsE} can be obtained by the multiplication of Eq.~(\ref{Faraday2}) by $B_i(\mathbf{k})$. The calculations are more straightforward in this situation. Below, we give the detailed derivation of this kinetic equation in order to show why the contribution of the turbulent term $\alpha_d$ entering the parameter $\alpha_+$ in Eq. (\ref{parameters}) is opposite in sign  to the case $\alpha_-$ in Eq.~(\ref{helicity-spectrum3}). Note that the parameter $\alpha_d(t)=\alpha_\mathrm{B}(t)$ enters the factor $\alpha_+$ with the opposite sign compared to the parameter $\alpha_\mathrm{B}(t)$ alone found in Ref.~\cite{Campanelli:2007tc} as well as in Eqs.~(37) and~(38) in Ref.~\cite{Pavlovich}.

Multiplying Eq.~(\ref{Faraday2}) by $B_i(\mathbf{k})$ and using the two-point correlator in Eq.~(\ref{correlator}), one obtains on the left-hand side
\begin{equation}\label{lhs}
  \frac{1}{2}
  \langle
    B^2(\mathbf{k},t)
  \rangle
  + \eta_m k^2
  \langle
    B^2(\mathbf{k},t)
  \rangle =
  \frac{(2\pi)^5\delta^{(3)}(2\mathbf{k})}{2k^2}
  [\partial_t\mathcal{E}_\mathrm{B}(k,t) + 2\eta_m k^2\mathcal{E}_\mathrm{B}(k,t)],
\end{equation}
where we use Eq.~\eqref{eq:EHSA}. On the right-hand side, multiplying Eq.~(\ref{Faraday2}) by $B_i(\mathbf{k},t)$ and using consistently the four-point correlator in Eq.~(\ref{four-point}) and then Eq.~(\ref{correlator}), one obtains the double integral
\begin{align}
  &\varepsilon_{ijk}k_j \frac{\tau_d}{P+\rho}
  \int \frac{\mathrm{d}^3p}{(2\pi)^3}
  \int \frac{\mathrm{d}^3q}{(2\pi)^3}\,q_r
  \nonumber
  \displaybreak[1]
  \\
  &
  \times \langle B_s(\mathbf{q})B_i(\mathbf{k})
  [\varepsilon_{krs}B_n(\mathbf{p} -\mathbf{q})B_n(\mathbf{k} - \mathbf{p}) -
  \varepsilon_{rsm}B_k(\mathbf{k} - \mathbf{p})B_m(\mathbf{p} -\mathbf{q})]\rangle
  \notag
  \displaybreak[1]
  \\
  & =
  \varepsilon_{ijk}k_j \frac{\tau_d}{4(P+\rho)}\int \mathrm{d}^3p\int \mathrm{d}^3q\,q_r
  \label{rhs-i}
  \displaybreak[1]
  \\
  & \times
  \Bigl[
  \varepsilon_{krs}
  \Bigl(
  \delta^{(3)}(\mathbf{q} + \mathbf{k})\delta^{(3)}({\mathbf{k}} - \mathbf{q})2S(p,t)
  [(\delta_{is} - \hat{k}_i\hat{k}_s])S(k,t) + \mathrm{i}\varepsilon_{ist}\hat{k}_tA(k,t)]
  \label{q+k,k-q}
  \displaybreak[1]
  \\
  & +
  \delta^{(3)}(\mathbf{p})\delta^{(3)}(2\mathbf{k} - \mathbf{p})
  [(\delta_{sn} - \hat{q}_s\hat{q}_n)S(q,t) + \mathrm{i}\varepsilon_{snt}\hat{q}_tA(q,t)]
  \label{p,k-p}
  \displaybreak[1]
  \\
  & \times
  [(\delta_{in} -\hat{k}_i\hat{k}_n)S(k,t) + \mathrm{i}\varepsilon_{inq}\hat{k}_qA(k,t)] +
  \delta^{(3)}(\mathbf{q} + \mathbf{k} - \mathbf{p})
  \delta^{(3)}({\mathbf{k}} + \mathbf{p} - \mathbf{q})
  \label{q+k-p,k+p-q,1}
  \displaybreak[1]
  \\
  & \times
  [(\delta_{sn} - \hat{q}_s\hat{q}_n)S(q,t) + \mathrm{i}\varepsilon_{snt}\hat{q}_tA(q,t)]
  [(\delta_{in} -\hat{k}_i\hat{k}_n)S(k,t) + \mathrm{i}\varepsilon_{inq}\hat{k}_qA(k,t)]
  \Bigr)
  \label{q+k-p,k+p-q,2}
  \displaybreak[1]
  \\
  & -
  \varepsilon_{rsm}\Bigl(\delta^{(3)}(\mathbf{q} + \mathbf{k})
  \delta^{(3)}({\mathbf{k}} - \mathbf{q})
  [(\delta_{is} - \hat{k}_i\hat{k}_s])S(k,t) + \mathrm{i}\varepsilon_{ist}\hat{k}_tA(k,t)]
  \label{q+k,k-q,1}
  \displaybreak[1]
  \\
  & \times
  [(\delta_{km} - \hat{p}_k\hat{p}_m)S(p,t) +\mathrm{i}\varepsilon_{kmq}\hat{p}_qA(p,t)]
  \label{q+k,k-q,2}
  \displaybreak[1]
  \\
  & +
  \delta^{(3)}(\mathbf{q} + \mathbf{k} - \mathbf{p})
  \delta^{(3)}({\mathbf{k}} + \mathbf{p} - \mathbf{q})
  [(\delta_{sk} - \hat{q}_s\hat{q}_k)S(q,t) + \mathrm{i}\varepsilon_{skt}\hat{q}_tA(q,t)]
  \label{q+k-p,k+p-q,12}
  \displaybreak[1]
  \\
  & \times
  [(\delta_{im} -\hat{k}_i\hat{k}_m)S(k,t) + \mathrm{i}\varepsilon_{imq}\hat{k}_qA(k,t)]
  \label{q+k-p,k+p-q,22}
  \displaybreak[1]
  \\
  & +
  \delta^{(3)}(\mathbf{p})\delta^{(3)}(2\mathbf{k} - \mathbf{p})
  [(\delta_{sm} - \hat{q}_s\hat{q}_m)S(q,t) + \mathrm{i}\varepsilon_{smt}\hat{q}_tA(q,t)]
  \label{p,2k-p,2}
  \displaybreak[1]
  \\
  & \times
  [(\delta_{ik} -\hat{k}_i\hat{k}_k)S(k,t) + \mathrm{i}\varepsilon_{ikq}\hat{k}_qA(k,t)]
  \Bigr)
  \Bigr].
  \label{rhs-f}
\end{align}
Let us list consistently the results of integration in the cumbersome Eqs.~(\ref{rhs-i})-\eqref{rhs-f}. The integrand in Eq.~\eqref{q+k,k-q} for terms $\sim \delta^{(3)}(\mathbf{q} + \mathbf{k})\delta^{(3)}({\mathbf{k}} - \mathbf{q})$ results in
\begin{equation}\label{third}
  -\frac{2(2\pi)^5\delta^{(3)}(2\mathbf{k})\tau_d}{P + \rho}
  E_\mathrm{B}(t)\mathcal{E}_\mathrm{B}(k,t),
\end{equation}
where $E_\mathrm{B}(t)=(2\pi)^{-2}\int_0^{\infty}p^2S(p,t)\mathrm{d}p$ is the magnetic energy density in the volume $V$ since, using Eq.~\eqref{eq:EHSA}, one gets that $E_\mathrm{B}(t)=(2V)^{-1}\int_V\mathrm{d}^3x\langle \mathbf{B}^2\rangle=\int_0^{\infty}\mathrm{d}k\mathcal{E}_\mathrm{B}(k,t)$. The integration of Eqs.~\eqref{p,k-p}-\eqref{q+k-p,k+p-q,2} using $\delta^{(3)}(\mathbf{p})\delta^{(3)}(2\mathbf{k} - \mathbf{p})$ and $\delta^{(3)}(\mathbf{q} + \mathbf{k} - \mathbf{p})\delta^{(3)}({\mathbf{k}} + \mathbf{p} - \mathbf{q})$ leads to the result
\begin{equation}\label{456}
  -\frac{(2\pi)^5\delta^{(3)}(2\mathbf{k})\tau_d}{3(P + \rho)}
  \frac{\dot{{H}}_\mathrm{B}(t)}{2\eta_m}\mathcal{H}_\mathrm{B}(k,t),
\end{equation}
where in standard MHD, $\dot{H}_\mathrm{B}(t)=- (2\eta_m/V)\int_V \mathrm{d}^3x\langle(\mathbf{J}\cdot\mathbf{B})\rangle =-2\eta_m\int_0^{\infty}\mathrm{d}kk^2\mathcal{H}_\mathrm{B}(k,t)$ is the temporal derivative of the magnetic helicity density. Note that Eqs.~(\ref{third}) and~(\ref{456}) result from the sum of the terms within the parentheses $\Bigl(...\Bigr)$ which is
proportional to the tensor $\varepsilon_{krs}$ in Eqs.~(\ref{rhs-i})--\eqref{rhs-f}, while from the next sum within the parentheses $\Bigl(...\Bigr)$ which is proportional to the tensor $\varepsilon_{rsm}$, one obtains in Eqs.~\eqref{q+k,k-q,1} and~\eqref{q+k,k-q,2} for the terms $\sim \delta^{(3)}(\mathbf{q} + \mathbf{k})\delta^{(3)}({\mathbf{k}} - \mathbf{q})$
\begin{equation}\label{78}
  \frac{2(2\pi)^5\delta^{(3)}(2\mathbf{k})\tau_d}{3(P + \rho)}
  E_\mathrm{B}(t)\mathcal{E}_\mathrm{B}(k,t),
\end{equation}
as well as in Eqs.~\eqref{q+k-p,k+p-q,12} and~\eqref{q+k-p,k+p-q,22} for the terms $\sim \delta^{(3)}(\mathbf{q} + \mathbf{k} - \mathbf{p})\delta^{(3)}({\mathbf{k}} + \mathbf{p} - \mathbf{q})$)
\begin{equation}\label{910}
  \frac{(2\pi)^5\delta^{(3)}(2\mathbf{k})\tau_d}{6(P + \rho)}
  \frac{\dot{{H}}_\mathrm{B}(t)}{2\eta_m}\mathcal{H}_\mathrm{B}(k,t).
\end{equation}
Finally, in Eqs.~\eqref{p,2k-p,2} and~\eqref{rhs-f}, for the terms $\sim \delta^{(3)}(\mathbf{p})\delta^{(3)}(2\mathbf{k} - \mathbf{p})$), one gets
\begin{equation}\label{1112}
  \frac{(2\pi)^5\delta^{(3)}(2\mathbf{k})\tau_d}{2(P + \rho)}
  \frac{\dot{{H}}_\mathrm{B}(t)}{2\eta_m}\mathcal{H}_\mathrm{B}(k,t).
\end{equation}
Summing the contributions in Eqs.~(\ref{third}) and~(\ref{78}), we obtain
\begin{equation}\label{diffusion_add}
  -\frac{4(2\pi)^5\delta^{(3)}(2\mathbf{k})\tau_d}{3(P + \rho)}
  E_\mathrm{B}(t)\mathcal{E}_\mathrm{B}(k,t),
\end{equation}
which, together with the magnetic diffusion parameter $\eta_m=(\sigma_\mathrm{cond})^{-1}$ in Eq.~(\ref{lhs}), gives $\eta_\mathrm{eff}=\eta_m + 4E_\mathrm{B}(t)\tau_d/3(P + \rho)$ coinciding with the result of Ref.~\cite{Campanelli:2007tc}, except for the factor $(P + \rho)$ in the denominator missed there; cf. Eq.~(\ref{parameters}) above.

Then, summing Eqs.~(\ref{456}), (\ref{910}), and~(\ref{1112}), we get the turbulence contribution to the evolution equation as
\begin{align}\label{turbulent_term}
  \frac{(2\pi)^5\delta^{(3)}(2\mathbf{k})\tau_d}{3(P + \rho)}
  \frac{\dot{{H}}_\mathrm{B}(t)}{2\eta_m} \mathcal{H}_\mathrm{B}(k,t) = &
  -\frac{(2\pi)^5\delta^{(3)}(2\mathbf{k})\tau_d}{3(P + \rho)}
  \mathcal{H}_\mathrm{B}(k,t)
  \int_0^{\infty}\mathrm{d}pp^2{\mathcal{H}}_\mathrm{B}(p,t)
  \notag
  \\
  & =
  -(2\pi)^5\delta^{(3)}(2\mathbf{k})\alpha_d \mathcal{H}_\mathrm{B}(k,t)/2,
\end{align}
where $\alpha_d$ is defined in Eq.~(\ref{parameters}), and we use the MHD relation as after Eq.~(\ref{456}).

Finally, combining Eqs.~(\ref{lhs}), (\ref{diffusion_add}), and~(\ref{turbulent_term}), separating the factor $(2\pi)^5\delta^{(3)}(2\mathbf{k})$ and multiplying both sides by $2k^2$, we reproduce Eq. (\ref{kineticsE}) for the energy spectrum evolution,
\begin{equation}\label{evolution_energy}
  \frac{\partial\mathcal{E}_\mathrm{B}(k,t)}{\partial t} =
  -2k^{2}\eta_\mathrm{eff}\mathcal{E}_\mathrm{B}(k,t) +
  \alpha_{+}k^{2}\mathcal{H}_\mathrm{B}(k,t),
\end{equation}
where the factor $\alpha_+=\alpha_\mathrm{CME} -\alpha_d$ includes the CME term $\alpha_\mathrm{CME}=2\alpha_\mathrm{em}\mu_5(t)/\pi\sigma_\mathrm{cond}$, as well as in the helicity spectrum evolution given by Eq.~(\ref{helicity-spectrum3}).

In spite of the coincidence of signs of the turbulent term $\alpha_d(t)$ and the analogous $\alpha_\mathrm{B}(t)$ in Refs.~\cite{Campanelli:2007tc,Pavlovich}, $\alpha_d(t)=\alpha_\mathrm{B}(t)= -\tau_d\dot{{H}}_\mathrm{B}(t)/3\eta_m(P+\rho)$, one can see that their contributions to the evolution equation for the energy density spectrum $\mathcal{E}_\mathrm{B}(k,t)$ are opposite when comparing our Eq.~(\ref{evolution_energy}) and, e.g., Eq.~(7) in Ref.~\cite{Campanelli:2007tc}.
Such a difference between our results and those in Refs.~\cite{Campanelli:2007tc,Pavlovich} exists only in the transport equation for $\mathcal{E}_\mathrm{B}(k,t)$, while kinetic equations for the helicity density spectrum coincide when $\mu_5=0$; see here Eq.~(\ref{helicity-spectrum3}).


\begin{thebibliography}{100}
\bibitem{fn1}
  The drag time is estimated as the free path
  $l_\mathrm{free}=(n\sigma)^{-1}\approx 1/\alpha_\mathrm{em}^2T$,
  where $n\sim T^3$ is the hot plasma density and
  $\sigma \sim \alpha_\mathrm{em}^2/T^2$ is the Coulomb cross section.
\bibitem{fn3}
  The mean binary product
  $\langle A_i(-\mathbf{k})B_i(\mathbf{k})\rangle =
  \mathrm{i}\varepsilon_{iqg}k_q\langle B_g(-\mathbf{k})B_i(\mathbf{k})\rangle/k^2$,
  where $A_i(-\mathbf{k})=J_i(-\mathbf{k})/k^2 =
  \mathrm{i}\varepsilon_{iqg}k_qB_g(-\mathbf{k})/k^2$,
  in accordance with the rule in Eq.~\eqref{correlator} equals to
  $(2\pi)^5\delta ^{(3)}(0)\mathcal{H}_\mathrm{B}(k,t)/2k^2$, where 
  $\mathcal{H}_\mathrm{B}(k,t)=kA(k,t)/2\pi^2$ is the helicity density spectrum.
\end{thebibliography}
\end{document}